\documentstyle[pre,aps,multicol,epsf]{revtex}

\newcommand{\mean}[1]{\left \langle #1 \right \rangle}
\newcommand{\beq}{\begin{equation}}
\newcommand{\eeq}{\end{equation}}
\newcommand{\beqn}{\begin{eqnarray}}
\newcommand{\eeqn}{\end{eqnarray}}

\newcommand{\dd}{\mbox{d}}

\renewcommand{\tanh}{ {\rm th}}
\renewcommand{\sinh}{ {\rm sh}}
\renewcommand{\cosh}{ {\rm ch}}

\newcommand{\includegraphics}{\epsfbox}
\newcommand{\resizebox}[3]{
\epsfxsize #1
\epsfysize #2
 #3
}

\begin{document}

\title{Activated Drift Motion of a Classical Particle With a Dynamical
Pinning Effect.} 
\author{Fabrice{\sc Thalmann}} 
\address{LEPES-CNRS {\it
BP166X  25 av, des Martyrs 38042 Grenoble Cedex} France\\
{\it thalmann@lepes.polycnrs-gre.fr}}
\date{June $29^{th}$ 1998}
\maketitle
\begin{abstract}
 A one dimensional trap model for a thermally activated
classical particle is introduced to si\-mulate driven dynamics in presence of
``ageing'' effects. The  depth of each  trap increases with the time elapsed
since the particle has fallen into it. The consequences  of this dynamical
pinning are studied, and velocity-force characteristics are  numerically
obtained. A special attention is paid to the situation where the particle is
pulled with a  spring to ensure a finite average velocity.  In the low velocity
regime, the presence of a broad distribution of trapping  times leads to
suppression of linear response, replaced by a threshold or by sublinear
dynamics. A regime of strong  fluctuations is obtained when the particle is
driven at intermediate velocities.
\end{abstract}
\pacs{05.40.+j Fluctuation phenomena, random processes,
and Brownian motion} 
%
%

\begin{multicols}{2}


\section{Introduction}
\label{sec:intro}

The out-of-equilibrium dynamics of glassy systems is a widely open subject. An
important topic concerns models for which a spontaneous ageing behavior
competes with an external field. This article aims to study a realization of
the above paradigm in a simple ``friction'' experiment where the  two main
features --ageing and external forcing-- are present. For this purpose, a 
stochastic process describing a classical particle in a one-dimensional pinning
potential is defined. The pinning potential consists of identical traps with
time-dependent depths. The barriers between two neighbouring traps  therefore
increase with time in order to simulate ageing  effects, and
the hopping rate between traps is based on Arrhenius dynamics. Some physical
justifications for such time-dependent barriers are exposed in section
\ref{sec:pinning}.  Indeed, by choosing adequately the time  dependence of the
trapping potential, one can account for anomalous slow diffusion situations
related to broad algebraic distributions of waiting times and leading to glassy
behaviors \cite{FeiVin,Bou}. One can alternatively  consider
a simpler case where the barriers increases from a short-time value to a
larger long-time value and discuss the related dynamical consequences.

Our main purpose is to understand the driven dyna\-mics of this
system in the situation where the average velocity $v$ is non zero. 
A quadratic potential moving at velocity $v$ is therefore added, acting in a way 
similar to a spring 
pulling a body in a friction experiment. The mean spring extension defines the 
friction force $F$.
The relation between $v$ and $F$ is called the ``Velocity-Force''
characteristics, and is the central output for such a friction experiment. Other
quantities of interest will be monitored like the distribution of waiting times
between hops or the histogram of the ``spring lengths''. In some situations, the
time-dependent pinning potential is found to enhance the fluctuations of the
particle position.

The basic feature of this model consists in a competition between the
``ageing'' pinning coming from the traps, and the renewal due to the driven
dynamics. It results in a strongly non-linear $v-F$ characteristics for a wide
intermediate velocity regime. The more extreme case of logarithmic growing
barriers leads to a non-ohmic behaviour at low velocity, typical of
glassy systems. In the latter case, a naive, ``mean-field like'', regularization 
with a
characteristic interruption time  $t_c(v)$,
fails in describing the low-velocity properties of the model. The limit $v \to
0$ indeed appears to be singular and involves large pinning times $\sim 1/v$.

Section II provides physical justifications of 
 the model, detailed in
Section \ref{sec:model}. A physical application is discussed. Section 
\ref{sec:ageing}
focuses on the velocity-force characteristics induced by logarithmic growing
barriers. Then, section \ref{sec:fluctuations} studies some consequences of
rapidly growing barriers on the particle's dynamics.


\section{A review of mechanisms generating an effective time dependent
pinning potential.}
\label{sec:pinning}

\paragraph*{A time dependent pinning potential.}

Many disordered systems like spin glasses, pinned random manifolds,
diffusing particles in random media, involve a disordered energy landscape. In 
such
systems, thermal effects, --diffusion and activation-- compete against pinning
effects. At low enough temperature, the out-of-equilibrium dynamics is usually
governed by activated barrier crossings in the configuration space. When the
distribution of energy barriers is broad, the time scale for reaching a thermal
--Boltzmann-- equilibrium may diverge with the size of the system. 

By following a particle during its out-of-equilibrium dynamics, one can attempt
to define a function of time, with dimension of an energy, as the mean (or
typical) height of the barrier that the system has to overcome, in order to
expand further. This function ``${\cal H}(t)$'' will be the  starting point for
this phenomenological approach of driven dynamics.

For instance, many aspects of the slow thermal diffusion of a particle in a 
one-dimensional quenched random force field, first introduced by Sina\"{\i} , 
have
been partially reinterpreted as the consequence of such effective barriers,
slowly increasing with time \cite{FeiVin}. The random force generates a pinning
potential --in 1 dimension-- with unbound local extrema, and the particle has
to overcome higher and higher barriers in trying to reach its thermal
equilibrium. Let $t_w$ be the time interval elapsed since the beginning of the
diffusion process, or waiting time. After $t_w$, the particle is shown to be at
equilibrium  within a restricted area bounded by a barrier of height ${\cal
H}(t_w) \sim T\ln(t_w)$ \cite{FeiVin,May}. Such a growing effective barrier
induces a strong dependence in $t_w$, or ageing, for the response and diffusion
properties of the particle, together with an ``anomalous'' slow diffusion
behavior. A generalization to other types of correlated disorder potentials 
leads to a variety of interesting regimes with sublinear characteristics 
\cite{Sch,LeDVin}.

In a related way, the ageing properties of  a particle diffusing in a high
dimensional phase space within traps with exponentially distributed depths,
were extensively studied, in relation with the magnetic relaxation of spin
glasses \cite{Bou}. Again, after a waiting time $t_w$, the particle stays in a
trap of typical depth ${\cal H} (t_w) \sim T \ln
(t_w)$ \cite{remarque2}.  Such a situation may occur for instance when 
describing the creep of an
extended object like an elastic manifold. The pinning energy of this object 
increases 
as its internal
degrees of freedom wander and find favorable configurations. One expects that
the longer the object stays at the same place, the larger its pinning energy
will be. A well known example involving pinning and
motion of elastic lines concerns the flux line problem in high Tc
superconductors \cite{BlaGesFeiLarVin}. 


 However, we must keep in mind that many other systems have
slow relaxation properties which do not originate from a barrier pinning mechanism.
This is the case for a particle evolving in a high dimensional,
random pinning potential \cite{rem:toymodel}.
In this system, thermal activated processes cannot explain the whole dynamics,
because a non trivial relaxation subsists when the temperature goes to 0.
In such situations, the slow dynamics is attributed to ``entropic effects'', 
{\it i.e} the system has more and  more difficulties to find favorable 
regions in its phase space, even if it is not separated from them by any 
energy barrier \cite{rem:entropic}.
The existence of growing effective energy barriers is therefore only a 
sufficient condition for ageing.


In what follows, Section IV deals with particles 
diffusing among traps whose depth increases
like $T\ln(t_w)$, leading to a power law distribution for waiting times. In the 
last section (\ref{sec:fluctuations}), we focus on a less singular case where 
the 
time-dependent barrier ${\cal H}(\infty) < \infty$ does not diverge.

\paragraph*{Detail of the friction experiment.}

In a friction experiment, one can either impose the external driving force
$\vec F$ or the velocity drift $\vec v$. In the first case, $\vec F$ is a
constant and one measures the mean position $\mean{\vec x(t)}$ of the particle. 
One can
alternatively pull the particle with a spring, with the other end moving at
constant velocity $v$. By doing so, one enforces only the average velocity, and 
the force
is then related to the mean extension of the spring. 

Some systems make these two procedures be inequivalent. Some examples of random
walks over random traps which exhibits only a sub-linear response for the mean
displacement $\mean{x(t)} \propto t^{\alpha} \ ; \ \alpha <1$ are reviewed
in\cite{BouGeo}. On the contrary, driving the particle with a spring ensures 
that
the average displacement $x(t)$ is  linear in time, in other words the average 
velocity always exists, 
provided the average is taken on sufficiently long times. Nevertheless, 
anomalous diffusion effects are
still present and appears through the effective friction force, with the 
disappearance of the
usual linear regime at low velocity. The study of such thermal ``creep'' for 
traps
models generalized in a way to allow the particle to be driven by a spring, will
be the main purpose of section IV. As far as possible, results for the spring
driven case will be compared to their ``constant force'' counterparts. 

On physical grounds, besides its ``regularization'' effect, the quadratic 
confinement potential may  reflect more complicated systems. For instance, when
describing a set of interacting particles, a crude mean-field  elastic
approximation  produces a ``cage potential'' (self-averaging contribution of
the whole system)  which limits the excursion of a given particle \cite{Fis:3}.
This  potential moves at velocity $v$ with the center of gravity of the entire
body. It can therefore be considered as  a first step towards the 
incorporation of interactions in anomalous diffusion problems.

As far as real ``dry'' friction experiments are concerned, it is known that 
phenomena similar to ageing occur at the contacts, and should be responsible
for  both the slow time dependence of threshold forces and for the
``stick-slip''  motion \cite{HesBauPerCarCar,CarNoz}. Spring pulling and
velocity dependent ageing are already present in Ref \cite{HesBauPerCarCar} but
their model is designed for a very different purpose. First, they introduce a
very strong ageing leading to a reentrant velocity-force characteristics and
unstable motion. Then, they consider only a crude ``mean-field'' approach,
called ``adiabatic approximation''.

In what follows we focus on overdamped,  thermally activated motion, and
especially on the low-velocity creep motion. In  contrast with the model of Ref
\cite{HesBauPerCarCar}, the ageing process is here described through a  broad
distribution of trapping times, and the dynamics is simulated, giving  access
to nonlinear creep characteristics as well as to fluctuation
effects. We provide details on the stochastic process and consider carefully
the whole distribution of waiting times and spring lengths, which turn out to
have a crucial importance for understanding the physical contents of this
model¬: the dangers of a naive mean-field approach are emphasized in
section IV.

\section{Trap dynamics in presence of an external driving force.}
\label{sec:model}

\begin{figure}[htbp]
   \resizebox{8cm}{5cm}{\includegraphics{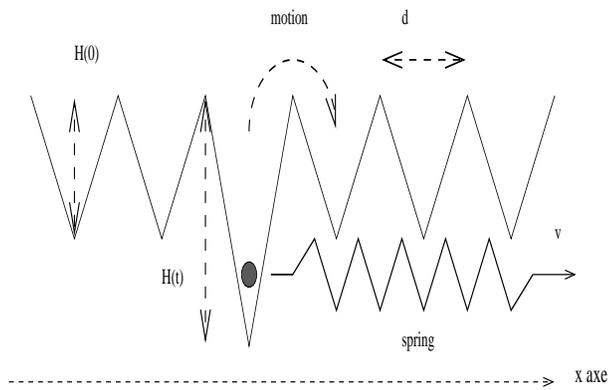}}
    \caption{Schematic illustration of the friction experiment.}
    \label{fig:potential}
\end{figure}

We consider a saw-tooth potential made of equidistant triangular wells (figure
\ref{fig:potential}). The interval between two sites is constant, equal to $d$.
The particle is supposed to hop towards its nearest neighbours sites, at a rate
given by the Arrhenius law. Whereas the conventional ``reaction rate theory''
requires additional information about the specific shape of the barrier, and
provides corrective terms \cite{HanTalBor}, we will keep only the crude
Arrhenius ratio. When a particle falls into a well, this one starts to
increase  its depth, in order to mimic an effective barrier ${\cal H}(t)$
growing with time. The corresponding escape rate decreases accordingly,
generating longer waiting times. A particle escaping from a well, immediately
falls into one of its  nearest neighbours. The particle is then always trapped,
excepted during the negligible transit delay over the barrier. The saw-tooth
potential is quite arbitrary, and the Arrhenius dynamics insensitive to this
specific shape.

\begin{figure}[htbp]
    \resizebox{8cm}{8cm}{\includegraphics{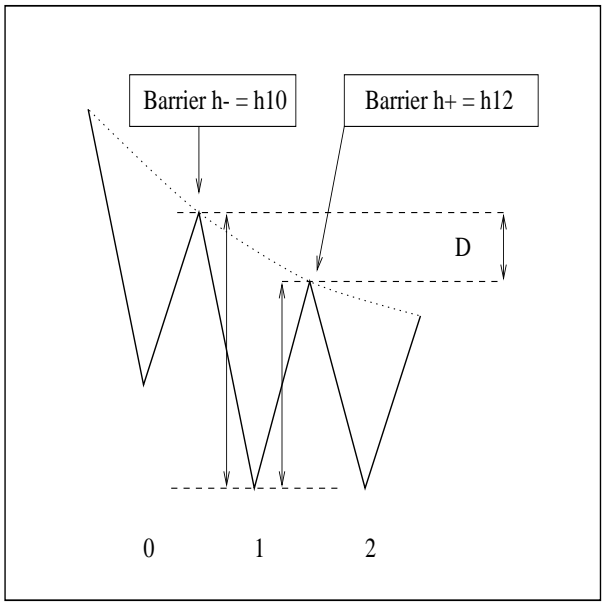} } 
     \caption{Barriers $h^+$ and $h^-$ (Section \protect\ref{sec:model}).}
    \label{fig:barriers}
\end{figure}

An external potential --constant driving force or spring traction-- tilts the
saw-tooth potential, reduces one of the two barriers and increases
the instantaneous escape rate. The barrier is strongly reduced by any
external bias, until it vanishes at the critical force $F_c(t)=2{\cal H}(t)/d$.
Let $h^-(\tau)$ and $h^+(\tau)$ be the height of the left and right barrier
(figure \ref{fig:barriers}).  $\tau$ is the time elapsed since the arrival
of the particle  in the occupied trap. By adding the left and right
``channels'', the total escape rate $w_{tot}$ reads, ($\beta$ standing for the
inverse temperature $T^{-1}$, $\omega_0$ for a trial frequency)~:

\beq
 w_{tot} (\tau) = \frac{\omega_0}{2} \exp (-\beta h^+(\tau)) + 
\frac{\omega_0}{2}
\exp (-\beta h^-(\tau)).
\label{eq:escaperate1}
\eeq

In absence of external force, the left and right barriers coincide and are set
equal to ${\cal H}(t)$. By choosing ${\cal H}(t)$ arbitrarily, one can
generate all the possible waiting time distributions.  ${\cal H}$ constant
corresponds to an ordinary activated diffusion over identical traps, leading to 
a
diffusion constant $D$ and mobility $\mu$:

\beq
 D =  \frac{\omega_0 d^2}{2} \exp(-\beta {\cal H}) \ , 
 \ \mu = \frac{\beta \omega_0 d^2}{2} \exp(-\beta {\cal H}).
\eeq

An external force $F$ changes $ {\cal H} \to \lbrace h^+ ={\cal H}  -Fd/2$; $h^-
={\cal H} +Fd/2 \rbrace $, leading to the constant force $F$ escape rate:

\beq 
w_{tot} (\tau) = \omega_0 \exp(-\beta {\cal H}(\tau)) \ 
\cosh \left( \frac{\beta F d}{2}
\right).
\label{eq:escaperate2}
\eeq

A spring of stiffness $k$, with a head
moving at constant velocity $v$ and the particle located at $x(t)$ exerts a
force $F(t)=kl(t)$ with $l(t)=vt-x(t)$. The total escape rate
\hbox{$w_{tot}(\tau,l_{(\tau=0)})$} reads~:

\beq
\omega_1 \exp[-\beta{\cal H}(\tau) + \beta{\cal H}(0)] 
\cdot  \cosh \left( \frac{\beta kd(vt-x(t))}{2} \right),
\label{eq:escaperate3}
\eeq
\[
\omega_1 = 
 \left\lbrace \omega_0 \exp \left( -\frac{\beta k d^2}{8} \right) 
\exp(-\beta {\cal H}(0)) \right\rbrace.  \nonumber
\]

It's essential to
distinguish between the total time $t$ and the delay $\tau$ relative to the
more recent jump of the particle, which occurred at $t-\tau$.  Once the value of
$l_{(\tau=0)}$ is known, at the time of arrival of the particle,
(\ref{eq:escaperate3}) determines the probability distribution for the waiting
time before the next jump.

As expected, the external force affects the escape rate. At large times, the
hyperbolic cosine dominates the exponential term and make the particle
eventually depin. Depinning is fast when $w_{tot}$ becomes larger than 1,
which happens when the force is greater than the critical ratio $ 2{\cal H}
(\tau )/d$.

Let's denote by $\Pi(t)$ the probability for the particle to stay within the
well for a time $\tau$ longer than $t$. $\Pi(t)$ decays exponentially as~:

\beq
  \Pi(t) =  \exp \left( - \int_0^{\,{\displaystyle t}}  
  \dd s\  w_{tot}(s,l_{(s=0)}) \right). 
\label{eq:Piwrelation}
\eeq

By definition, \noindent $\pi(t) = -\dd \Pi(t) / \dd t$ is the probability 
distribution of
waiting times. The knowledge of $\Pi(t)$ allows to perform ``molecular
dynamics'' simulations by following a procedure close to BKL \cite{BorKalLeb}.
In this simulation scheme, one computes at each step, the waiting time $\tau$
before the next jump. This is achieved by converting a uniform random number $r
\in [0,1]$ into $\tau(r)$, {\it i.e.} by inverting $\Pi(\tau)=r$.

The stochastic process consists in a sequence of jumps, labeled by $i$. Each
jump has a random direction \hbox{$\sigma_i=+/-1$} (left or right), and a random
waiting time $\tau_i~\in~\lbrack 0,\infty \lbrack$. The total time $t_n$, particle
 position $x_n$, and the spring extension $l_n$ follow the recurrence equations~:

\beq
\left\lbrace
\begin{array}{ccl}
t_{n} & = & t_{n-1} + \tau_n \\
x_{n} & = & x_{n-1} + \sigma_n d\\
l_{n} & = & l_{n-1} - \sigma_n d + v \tau_n
\end{array}
\right.
\label{eq:recurrence}
\eeq

A fundamental difference distinguishes constant force and  spring friction
experiments. In the first case, the waiting times $\tau_i$ are statistically
independent, while in the second case, they are strongly correlated, because
their probability distribution is conditioned by the spring extension $l(t)$,
which varies slowly. For example, a long waiting time extends the spring, and
forces the next waiting times to be shorter than the average. These
correlations prevent from performing a complete analytical study, and justify
a numerical approach. By defining
%
$V = (\beta kdv)/2$, $X(t) = (\beta kd x(t))/2$, 
$ L(t) = X(t) - V t $ , $\Delta = (\beta k d^2 )/2$,
the cosine term becomes $\cosh(Vt+L(t))$. A spring length $L(t) \sim
1$ means that the external force has reached its critical value $F_c \simeq
2{\cal H}/d $. $L(t)$ is piece-wise linear, increasing during the pinning of
the particle, and with discontinuities $-\sigma_i \Delta$ at each jump.  A
small value of $\Delta$ means that the particle has many different traps
accessible. At the opposite, for $\Delta \geq 1$, the spring forces the
particle to occupy only one given site or its nearest neighbours. This regime
has not  been considered here.

Provided the temperature is not too low, $\omega_0^{\! -1}$, $\omega_1^{\!
-1}$,  and the caracteristic time for the effective barrier are of the same
order and constitute a unique, microscopic, time scale. We have restrict
ourselves to this situation ($\beta \sim 1$). The only independent
parameters of the model turn to be~: those defining ${\cal H}$, the rescaled jump
length $\Delta$ and the ``velocity'' $V \omega_1^{\! -1}$. 

The mean friction force is given by

\beq
\mean{F} = \frac{2T}{d} \mean{L(t)},
\eeq

\noindent where brackets stand for the time average. \\
$ \mean{O} = \lim_{\Theta \to \infty}
\frac{1}{\Theta} \int_{0}^{\Theta} O(s) \, ds $.
The exact formulas for the mean position $L$ and its quadratic fluctuation
$L^2$ are~:

\beqn
\mean{L} & = & \lim_{n \to \infty} \frac{ \sum_{i=1}^{n} L(t_i) \tau_i + 
V \tau_i^2 /2}{ \sum_{i=1}^{n} \tau_i}. \nonumber \\
\mean{L^2} & = & \lim_{n \to \infty} \frac{ \sum_{i=1}^{n} L(t_i)^2 \tau_i 
+ L(t_i)V \tau_i^2 + V^2 \tau_i^3 /3 }{\sum_{i=1}^{n} \tau_i}. \nonumber 
\eeqn


There is no need to average over a quenched disorder in this model, and simulations
are notably simplified. We stress however that the  time averages mentioned above 
may present important fluctuations from one realization of the process to another.
This is especially the case when a broad distribution of trapping time is considered, 
and the displacement $x(t)$ as a function of time may exhibit fluctuations as  large 
as $\mean{x(t)}$ itself.
Note that our numerical work corresponds to spring pulling 
experiments, thus trapping time distributions  are regularized. We 
have checked carefully that the time averages were well defined and converged.


At each jump, the direction $\sigma_n=1$ is chosen
 with proba\-bility
$p(\tau_n,t_{n-1})$ and $\sigma_n=-1$ with probability \hbox{$1-p$}.
We made the choice

\beqn
p(\tau_n,t_{n-1}) & = & \frac{\exp (-\beta h^+(\tau_n))}
{\exp (-\beta h^+(\tau_n))+\exp (-\beta h^-(\tau_n))} \\
  & = &\! \! \frac{1}{1 + \exp [-2L(t_{n-1})-
  2V\cdot\tau_n] }.
\nonumber
\eeqn

Restricted to the case $V=0$ and ${\cal H}$ constant, the above choice fulfills
the detailed balance equation, and the resulting Boltzmann equilibrium
distribution for $l(t)$ is a gaussian with variance $T/k$.

One can demonstrate that the general case $V=0$, but ${\cal H}$
increasing, still leads to a stationary distribution profile for
$l(t)$, provided the mean trapping time exists, {\it i.e.} ${\cal H}$
must be a bounded or slowly increasing function of time.


\section{Large algebraic waiting time distributions.}
\label{sec:ageing}

\paragraph*{Definitions.}

It is well known that ageing effects appear when the waiting time distribution
decays like $\pi(\tau) \sim \tau^{-(1+\alpha)}$ and $\alpha
< 1$. The following choice ($\tau_1= 1/\omega_1$ is the reference time scale):

\beq
\Pi(t) = \left( 1+\frac{\tau}{\alpha \tau_1} \right)^{-\alpha} \ , \
\pi(t) = \frac{1}{\tau_1} \left( 1+\frac{\tau}{\alpha \tau_1}
\right)^{-(\alpha+1)},
\eeq
is obtained, following (\ref{eq:Piwrelation}), with

\beqn
\beta {\cal H}(t) & = & \ln \left( 1+\frac{\tau}{\alpha \tau_1} \right),
 \ \mbox{and} \\
w_{tot}(\tau) & = & \frac{\alpha}{\alpha \tau_1 + \tau}.
\label{eq:logincreasing}
\eeqn

When a constant force $F$ is applied, one has~:

\beq
w_{tot}(\, \tau \,) = \left( \frac{\alpha}{\alpha \tau_1 + \tau}
\right) \cdot  \cosh (\beta Fd/2).
\eeq

The effective exponent is no longer $\alpha$, but $\alpha'(F)=\alpha \
\cosh(\beta Fd/2)$ and the hopping length $d$ remains constant. This picture is
different from the one emerging from the Sina\"{\i} model, where the force
dependent exponent is $\alpha(F) = F/F_c$ ($F_c$ is a given critical force), and
the typical hopping length depends on F like $d(F) \sim 1/F^2$. This latter
phenomenology could be used as another starting point instead of equations
(\ref{eq:escaperate2},\ref{eq:escaperate3},\ref{eq:recurrence}).

The behaviour of a system with ${\cal H}$ growing as above, and a weak constant
force $F$  is just a particular case of the ``continuous time random walks''
described in \cite{BouGeo}. The mean displacement $\mean{x(t)}$ of the
particle turns out to be in this case (disregarding prefactors):

\beqn
\mean{x(t)} & = & N(t) \cdot d \cdot \tanh \left(
\frac{Fd}{2T} \right).  \label{eq:BouGeoDouresult1}\\
\mean{x(t)} & \simeq & \frac{d^{\, 2}}{2T} \cdot
\left( \frac{t}{\tau_1} \right)^{\alpha} \cdot F 
\label{eq:BouGeoDouresult2}
\eeqn

N(t) is the typical number of jumps during the time interval $[0,t]$, and
functions of $\beta F d/2$ have been linearized in the last equation. The
sublinear dependence in time means that the velocity is asymptotically $0$.

Let us turn now to the system with a spring pulling the particle at constant
velocity $v$. The constant force result (\ref{eq:BouGeoDouresult2}) is helpless
in this case. This is precisely a situation where driving with a spring is
inequivalent to driving with a constant force. In this case, the broad waiting 
time distribution is cut when the spring
depins the particle. A cut-off time $t_c$ is defined as~:

\beq
k l(t_c) = k v t_c = F_c = \frac{{\cal H}(t_c)}{d/2}, \nonumber
\eeq

\beq 
\frac{t_c}{ \ln \left( 1 + t_c/\alpha \tau_1 \right)}
= \frac{2}{k \beta d } \cdot \frac{1}{v}. 
\label{eq:cutofftime}
\eeq

\paragraph*{A mean field approach.}

One can try to understand the creep behaviour of the system by a mean field
argument, by neglecting all the fluctuations in the force and writing
$F=k \mean{l(t)}$. Owing to the regularization provided by the  velocity, the 
mean
value of the waiting times exists and equals:

\beq
\mean{ \tau(v) } = \tau_1 \left( \frac{t_c(v)}{\tau_1} \right)^{1-\alpha}.
\eeq

Result (\ref{eq:BouGeoDouresult1}) holds, and one can write a self-consistent
equation for $v$, $v \ll 1$.

\beq
v \cdot t = F \cdot \frac{d^2}{2T} \cdot \frac{1}{t_c(v)} \cdot \left(
\frac{t_c(v)}{\tau_1} \right)^{\alpha} \cdot t,
\eeq

\noindent  which, apart from a logarithmic factor, leads to $t_c(v)~\sim~1/v$
and~:

\beq 
F = \frac{2\tau_1}{\beta d^2} \left( \frac{2}{\beta k d \tau_1} 
  \right)^{1-\alpha} \cdot v^{\alpha}.
\label{eq:meanfieldresult}
\eeq

The linearized case $\beta F d/2 \le 1$ implies  $\alpha'= $ \\ \hbox{$\alpha \ \cosh(\beta
F d/2) \simeq \alpha $} (at weak velocity, the force does not change much the
value of $\alpha'$). This mean field approach gives a power law creep $F \sim
v^{\alpha}$. Unfortunately, it strongly disagrees with numerical simulations.

The simulations were done, following the scheme of equation 
(\ref{eq:recurrence}), section II. Details on the computational sche\-me are
deferred to the  Appendix~: the main difficulty comes
 from the conversion of random numbers in waiting times, and a trick is 
necessary to avoid computer time wasting.

\paragraph*{Analysis of the numerical results.}

\begin{figure}[htbp]
   \resizebox{8cm}{8cm} {\includegraphics{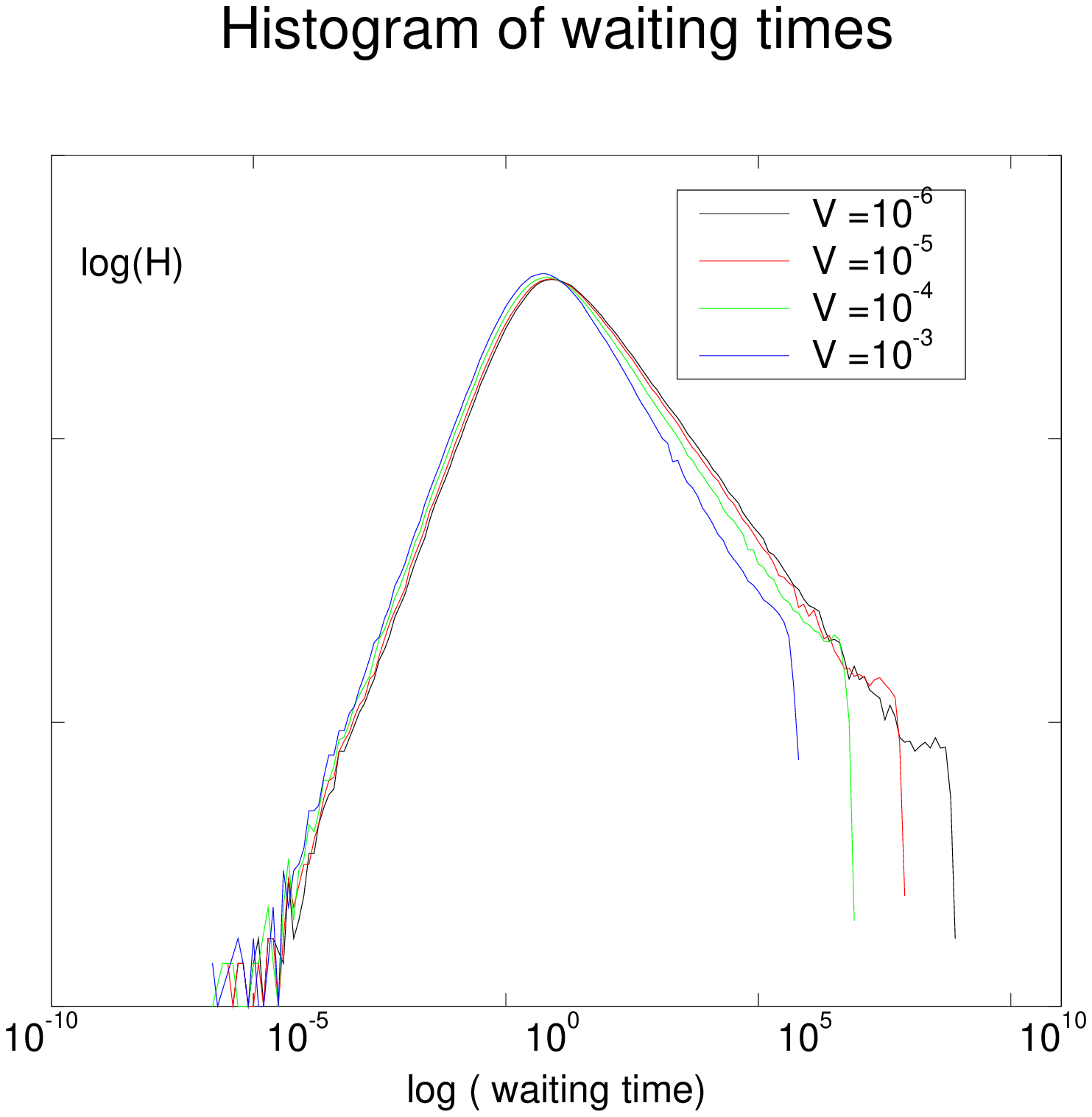}} 
   \caption{ Distribution of
     waiting times regularized by the velocity. ($\alpha=0.5$).  This
     curve $H(\ln t)$ has to be understood as follows~: the number of
     waiting times lying between $t_1$ and $t_2$ is $ N(t_1,t_2) =
     {\cal N} \int_{t_1}^{t_2} H(\ln t) d \ln t $. (${\cal N}$ is a
     normalization factor). $H(\ln t) \sim t^{-\alpha}$. }
     \label{fig:h_wt_alg.vel} 
\end{figure}

Figure (\ref{fig:h_wt_alg.vel}) shows an histogram for the distribution of
waiting times with $\alpha=0.5$, at various velocities. The distributions are
clearly cut at $t_c(v)$. The constant slope $-(1+\alpha')$ in the log-log plot
shows that one has true power laws until the cut-off. At intermediate 
velocities, $\alpha'$ departs from its $0$ velocity value, as expected.

\begin{figure}[htbp]
     \resizebox{8cm}{8cm} {\includegraphics{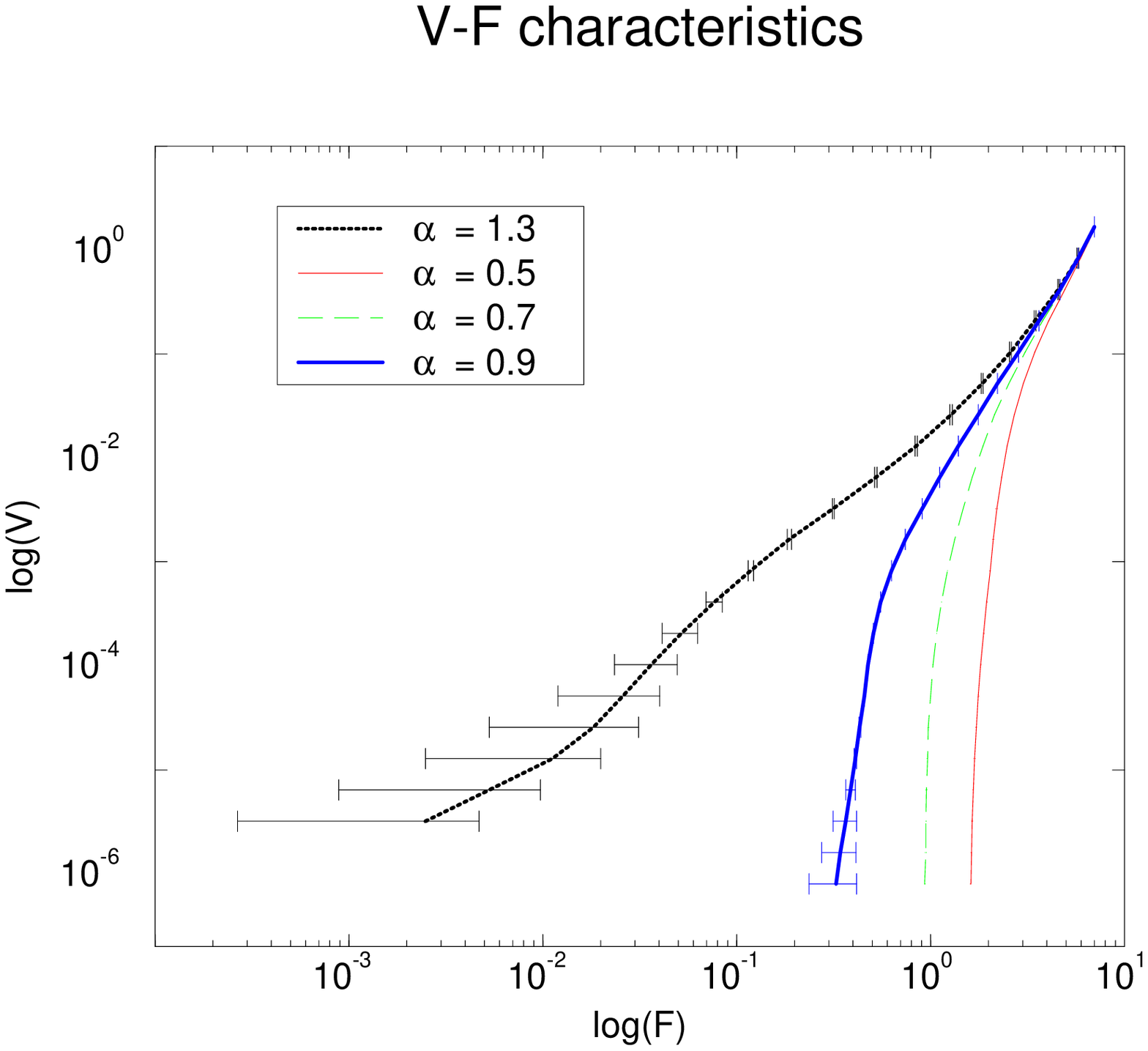} } 
     \caption{Dependence of the v-F curve in the exponent $\alpha$. The closer
    to 1 is $\alpha$, the more difficult is the convergence at low velocities.
  $\Delta = 0.005$; $V = 0.5 \beta k d $ ranges from $4.10^{-8}$ to $0.1$} 
     \label{fig:vf_alg.alpha}
\end{figure}

Figure (\ref{fig:vf_alg.alpha}) shows the dependence of the $v-F$ curve in the
exponent $\alpha$. The characteristics have been found to be monotonic and
increasing. For both $\alpha=0.9$ and $\alpha=1.3$, the curves are
presented with horizontal error bars (which are very small for $\alpha=0.5$
and $\alpha=0.7$). They give an estimate for the fluctuations between
successive runs. The convergence is all the more difficult to get that the
velocity is weak and the exponent $\alpha$ close to 1. All the characteristics
with $\alpha < 1$ have a threshold value decreasing while $\alpha$ becomes
closer to 1. For $\alpha > 1$, one should recover an ohmic regime at low
velocities. The characteristics for $\alpha = 1.3$ are compatible with such an
ohmic response, but fluctuations from one run to another remain important.
This is probably a consequence of the infinite mean squared value of the
waiting times when $v \to 0$

In log-log coordinates, the $v-F$ curves are concave, and seems at first glance
to end at a threshold value \hbox{$F(v\to 0)$} whereas in linear  coordinates, at the
opposite, they are convex.  Among the curves known to change their convexity
when the axes are logarithmically graduated, one finds for example the
celebrated stretched exponentials $v= v_0 \exp [-(f_c/f)^\gamma]$ occurring in
some glassy creep cases \cite{BlaGesFeiLarVin}. For $\alpha=0.5$ such a
functional form fits well the intermediate regime $10^{-4} < v < 10^{-1}$ with
$\gamma \sim 2.8 \pm 0.1$, but does not describe anymore the slower regime
 $v < 10^{-4}$.
This latter case is better fitted by a threshold characteristics $F_c +
(F-F_c)^{\beta}$, $\beta= 2.1$ (over only 1 decade).

One can roughly say that the $v-F$ curve, for $\alpha=0.5$, has a threshold
value $F_c$  with a force $F$ which does not vary by more than 12 \% over 2
decades. This is enough to discard the mean field description mentioned above 
\hbox{(eq \ref{eq:meanfieldresult})} . The other values of $\alpha=0.7;0.9$ exhibits a
similar threshold force, but as $\alpha$ becomes closer to 1, it is more and
more difficult to get a good, well converged value for the force at weak
velocity and the fit with the stretched exponential becomes poor. 

\begin{figure}[htbp]
     \resizebox{8cm}{8cm} {\includegraphics{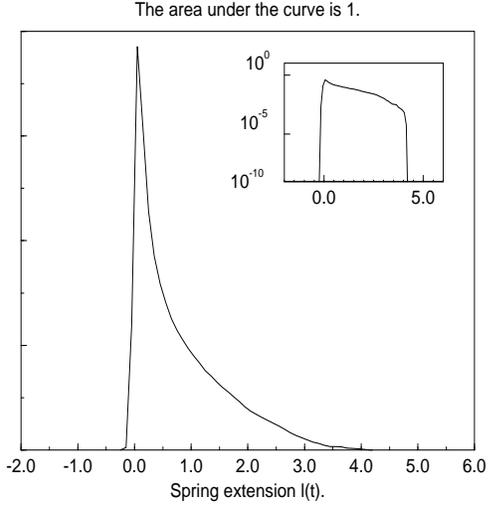} } 
      \caption{The histogram of spring lengths $L(t) \left( =\beta k d l(t)/2
    \right) $ does not center around 0,
   even for a vanishing velocity (here $v=10^{-6}$). The inset shows the
   histogram with a logarithmic vertical axe.}
      \label{fig:h_al_alg}
\end{figure}

One can account for the existence of the quasi-threshold behaviour of the $v-F$
curve ({\it i.e.} up to logarithmic factors) with a simple argument. The
distribution of waiting times is a broad power law distribution cut for times
greater than $\sim d/v$. The contribution of large waiting times $\tau \sim
d/v$ represents a finite ratio of the total sum $t = \sum_{i=1}^{n} \tau_i$.
Such waiting times correspond to a finite value for the spring extension $k v
\tau$, and occur at a constant rate. As they are not balanced by any negative
contribution, the resulting mean spring extension remains finite. Figure
(\ref{fig:h_al_alg}) confirms this scenario. The histogram for the variable
$l_i$ remains asymmetric, even for a vanishing velocity, leading to a constant
friction force. The ``mean-field'' approach fails because it relies on the
existence of a typical value $F$ for the force, vanishing with $v$, in
contradiction with figure (\ref{fig:h_al_alg}).

It is interesting to notice that the mean-field procedure leads to a (wrong)
power law creep at low velocity. Such power laws have been found by Horner in
the mean-field theory of a particle in a short range correlated disorder
\cite{Hor}. Whereas it is a rather different problem, where ageing comes from
slow relaxation in a high dimensional phase space, and not from any barrier
mechanism (section \ref{sec:pinning}), we emphasize that both
mean-field approaches lead to a power law  for $v \to 0$. One can suspect the
corrective terms to the above mean-field theory (which contain all thermal
activation effects) to modify the algebraic creep into a threshold, like
here. 

Anyway, in every physical case, it must exist a huge but finite equilibration
time $t_{eq}$, at which the effective barrier ${\cal H}$ stops increasing. The
above threshold characteristics must match with a slow thermally activated
ohmic regime for velocities $ v \leq d/t_{eq}$.


\section{Velocity induced fluctuations.}
\label{sec:fluctuations}

\paragraph* {A stepwise potential.}

The necessary and sufficient condition for the existence of an ohmic regime at
low velocity is that the mean trapping time $\mean{\tau}$ remains finite. This
is because the effective ``viscosity'' $\eta=\lim_{v \to 0} F/v$ scales like
$\mean{\tau}$. At the opposite, $\mean{\tau} = \infty$
implies  a sublinear or ``creep'' regime in the $v-F$ characteristics. This
latter can always be written¬:

\beq
F = \eta \cdot v,
\eeq

\noindent where $\eta$ depends on $F$ or on $v$ according to the friction
experiment considered. When using a time-dependent potential ${\cal H}$, one
can generate a distribution of waiting times strongly dependent on $v$, with
consequently, a non constant mean pinning time $\mean{\tau (v)}$, decreasing
with $v$. 

Whenever ${\cal H}$ remains bounded,  one expects a crossover between the
strong velocity regime involving ${\cal H}(\tau \simeq 0)$ and the low velocity
regime depending on the whole range $ \tau \in \lbrack 0, \infty \lbrack $.
Moreover, a detailed study of the fluctuations $\mean{l^2(t)}$ of the
particle's position turns out to be very instructive and reveals some
unexpected features.

In order to make apparent the effects of a non-constant ${\cal H}$ function, I
have introduced a decreasing step function. Such a sharp behaviour for ${\cal
H}$ should make the crossover well contrasted. This is done with the
escape rate~:

\beq
w^0_{tot} = 
\left\lbrace
\begin{array}{cr}
\omega_1 & {\rm for \ } \tau \in \lbrack 0, t_s \rbrack  \\
\omega_2 & {\rm for \ } \tau \in \lbrack t_s, \infty \lbrack 
\end{array},
\right.
\eeq

where $t_s$ is the arbitrary position of the step. The $\Pi(t)$ function
(\ref{eq:Piwrelation}) and its
reciprocal $\Pi^{-1}$ can be computed exactly, providing an easy numerical
generation of waiting times. $\Pi(t)$ equals~:

\beqn
t < t_s & : &  
\exp \left[ {\displaystyle \frac{\omega_1}{V} 
( \sinh(L) - \sinh(L+Vt))} \right], \nonumber\\
 t > t_s & : & 
\exp \left[ {\displaystyle \frac{\omega_1}{V}\sinh(L)
 + \frac{\omega_2 - \omega_1}{V}\  \sinh(L + V t_s)} \right. \nonumber\\
& & \left. {\displaystyle - \frac{\omega_2}{V}\sinh (L+Vt)} \right],
\eeqn

\paragraph*{The enhancement of fluctuations.}

The mean square value is computed and compared with the equilibrium one, given
by the energy equipartition theorem $k \mean{l^2(t)}= T$, where  $k$
denotes the spring stiffness.

\begin{figure}[htbp] 
    \resizebox{8cm}{8cm} {\includegraphics{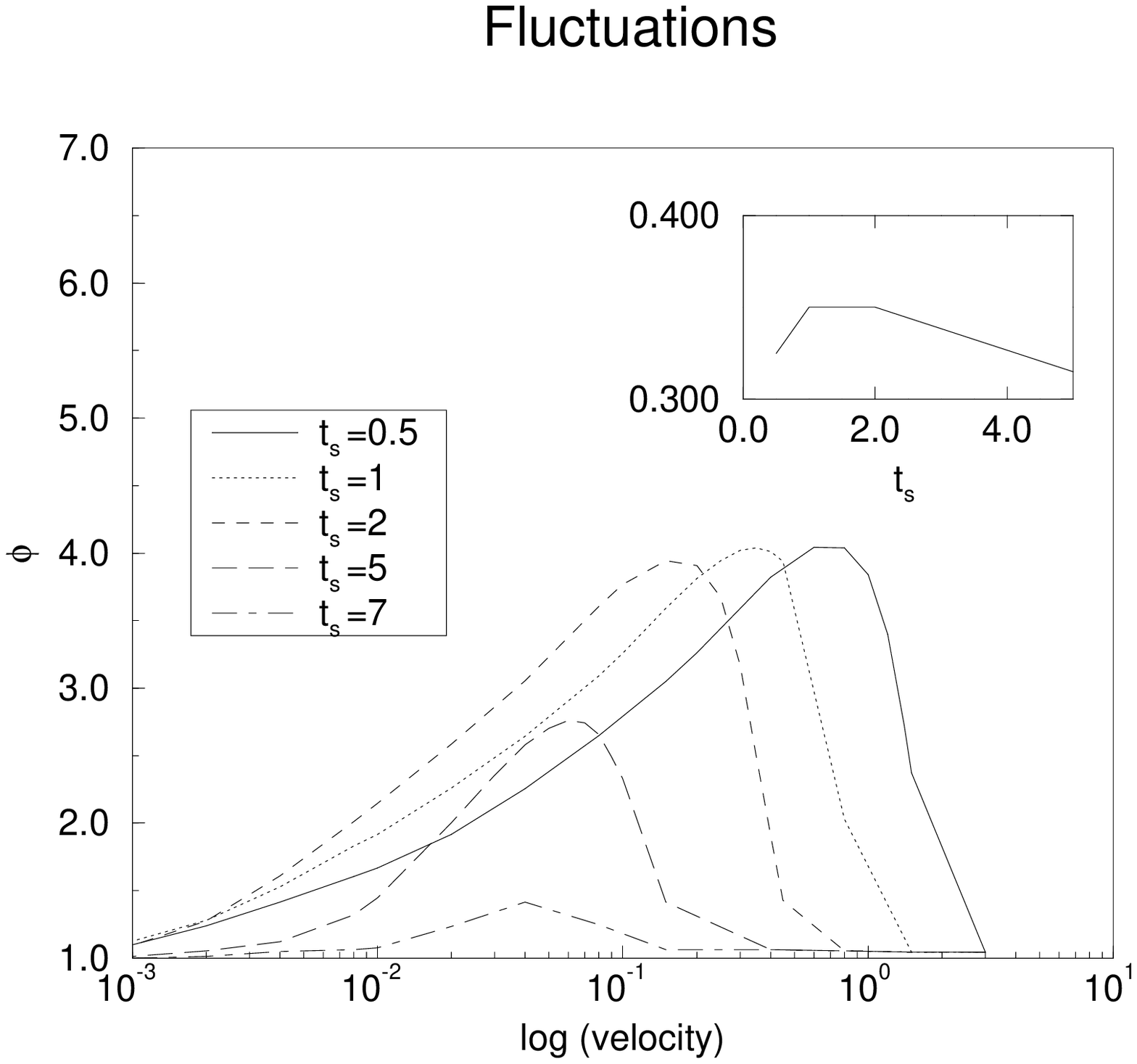} }    
    \caption{$\phi = k \beta \mean{l^2(t)}$ ratio between fluctuations and
 equilibrium fluctuations. The stepwise ${\cal H}$ cases S lead  to a maximum
 at $v=v_{max}$. The constant ${\cal H}$ cases A,B are characterized by $\phi
 \simeq 1$ (not on this picture). The inset shows the product  $v_{{\rm max}}
 \cdot t_s$ vs $t_s$, which has a constant value $\sim 0.33$. } 
    \label{fig:vf_fluct} 
\end{figure}

In figure (\ref{fig:vf_fluct}) the ratio $ \varphi = k\mean{l^2(t)} /T$ is 
plotted for the special choices $\omega_1=1$, $\omega_2=0.01 (= 1 \% \ 
\omega_1)$, 
$t_s$ ranging from 0.5 to 7. The results are compared with the constant ${\cal 
H}$ cases A: $w_{tot}^0 =
\omega_1$ and B: $w_{tot}^0 = \omega_2$. The step function case S interpolates
between these two static pinning potential cases. Thus, one expects the results
for S to fall in between the two extreme cases A and B. 

\begin{figure}[htbp]
     \resizebox{8cm}{8cm}{\includegraphics{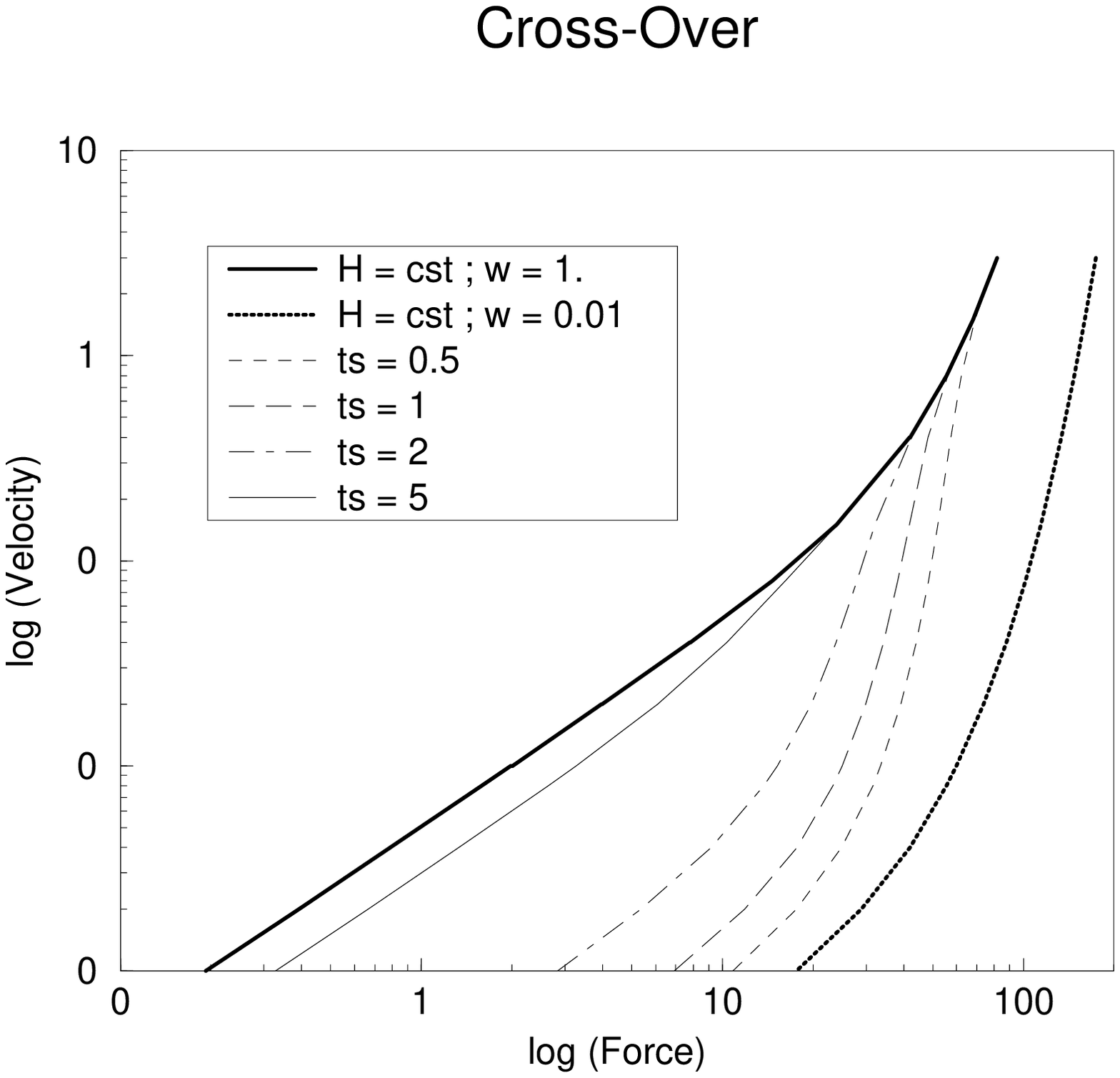} }     
    \caption{Comparison of the velocity force characteristics, between the
    constant $w$ cases A and B, and the stepwise $w$ cases S.
    The time $t_s$ of the step ranges from 0.5 to 5.}
    \label{fig:vfcrossover}
\end{figure}

The $v-F$ characteristics behaves in this way (see figure
(\ref{fig:vfcrossover})).
Whereas the constant $w_{tot}^0$ cases A and B exhibits fluctuations very close
to equilibrium ones, regardless to the value $\omega$, the stepwise situation S
shows a maximum for an intermediate regime  centered around $t_s^{-1}$.
Fluctuations are enhanced with a factor 3 or 4, before reaching their near
equilibrium value at high velocities (figure \ref{fig:vf_fluct}).

The value $\varphi \simeq 1$ at high velocities suggests a decoupling between
translation motion and fluctuations, similar to the one due to the galilean
invariance, in the disorderless case.  
This is reminiscent from a well known result about a fast particle moving on a
random potential. When considering a particle driven at high velocity, with a
force greater than the critical one, the pinning potential plays only a
perturbative role. The random pinning force acts like a Langevin noise and
leads to an effective ``shaking temperature'' scaling like the inverse 
velocity
$1/v$ (as far as a single particle is considered) \cite{reviewmovingphase}.

The effect of fluctuations is thus maximal at intermediate velocities. In the 
context of a friction experiment, such anomalous fluctuations are reminiscent
 of 
``stick-slip'' phenomena. On the other hand, in the physical case where the 
spring potential mimics the elastic interactions
between particles, a strong fluctuation regime indicates that the
motion occurs with strong deformations, with a possible breakdown of the 
elastic regime towards a plastic one. At the opposite, at weak or large
velocities, the motion of the particles may be coherent, within an homogeneous
flow \cite{reviewmovingphase} (two neighbouring particles remain close for a
long time).  The onset of such a ``plastic flow'' should occur when the amount
of fluctuations $\mean{x^2}$ reaches a critical value like in the celebrated
Lindemann criterion for melting.

\begin{figure}[htbp]
   \resizebox{8cm}{8cm} {\includegraphics{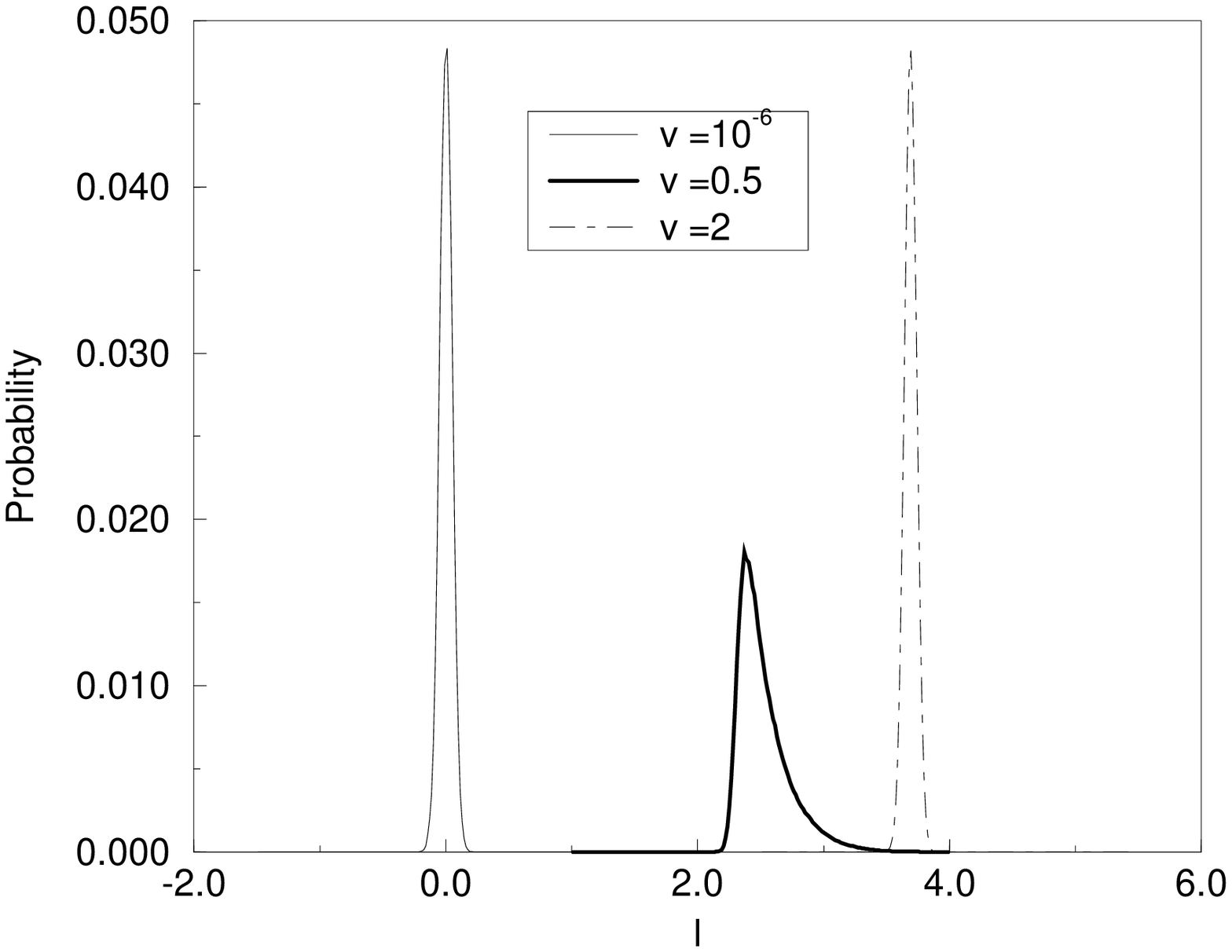} }  
    \caption{The distribution
     of spring lengths $L(t)$ is a gaussian at low and high velocities
     and is stretched at intermediate velocities.}
     \label{fig:h_al_step.vel} 
\end{figure}

Figure(\ref{fig:h_al_step.vel}) shows histograms for weak $v=10^{-5}$,
intermediate $v=0.5$ and high velocity  $v=2$. The step occurs at $t_s=1$.
$\varphi \simeq 1$ corresponds to a gaussian shape of the $l(t)$ distribution, 
$\varphi > 1$ coincides with an asymmetric distribution for $l(t)$, with a tail
on the right, related to large values of $l$. The presence of such strong
fluctuations requires a high enough velocity, able to ``convert'' a long
pinning time into a large value of $l$. At the other extreme, if $v$ is too
large, the friction force is higher than the critical force, and the pinning is
no more efficient. The effect is maximal for $ 0.32 < v t_s <0.35$.

\section{Conclusion}
\label{sec:conclusion}

In this paper, a general stochastic process has been defined, which allows 
dynamical pinning effects,
by introducing time-dependent barriers. The model has first been specialized to
generate power-law distributed waiting times. The case of a particle driven by
a spring has been investigated and compared to the previously known constant
driving force case. The sublinear drift of the latter case is found to have a 
counterpart which in the present study appears as
a finite threshold for the friction force, even for vanishing velocities. A 
naive mean-field approach fails in describing the velocity-force 
charac\-teristics, showing the prominent role of large waiting times, even 
within a distribution cut-off by the drift motion. 
The analogy with the $D=\infty$ ``mean field'' result suggests that the finite
dimension corrections could drive the  power-law creep into a 
threshold characteristics, ultimately linear at very low velocities if the 
distribution of waiting times is cut by size effects.

Then, we have investigated the consequences of a time dependent pinning
potential concerning the fluctuations in the particle position. It has been
achieved with a sharp step profile for the time dependent barrier. Fluctuations
are strongly enhanced at an intermediate velocity regime. This phenomenon is 
important in various physical cases where the simple model presented here 
applies and deserves more investigations.

More generally, this model provides a general framework in order to test the
equivalence between applied force and applied velocity in a friction
experiment, with presence of ``surface ageing'' and  at a scale where the
temperature is a relevant parameter.

\acknowledgements         
I thank J.P Bouchaud, A.Valat, J. Farago, J-L. Gilson and D. Feinberg 
for fruitful discussions,
and D. Feinberg especially for a critical reading of the manuscript. I am
indebted to W. Krauth for his lecture about Monte-Carlo methods at Beg Rohu
Summer School of Physics (France) 1996.


\section*{Appendix:Random generation of waiting times.}
\label{app:numericalscheme}

Given a random number $r$ taken from a uniform distribution over $[0,1]$, the
corres\-ponding waiting time $\tau$ is $\Pi^{-1}(r)$, with $\Pi^{-1}(\Pi(s))=s$
and $\Pi$ defined by (\ref{eq:Piwrelation}).  As the integration cannot be --in
the general case-- performed analytically, a direct evaluation of $\tau$
requires to find numerically the root of a function defined by a quadrature.
Independently, convergence requires many millions ($10^7$ to $10^8$) numbers
$\tau_i$ at low velocities, excluding any direct computation scheme for
$\Pi^{-1}$.

It's natural to look for asymptotic approximations of the integral. First, one
notices that the integrand \\ \hbox{$w_{tot}^{0}(\tau) \cosh(Vt +L)$} has at least two 
very
different time scales $\omega_1^{-1}$ and $V^{-1}$.

Let us detail the particular case $\alpha\cosh(Vt+L)/(\alpha \tau_1 + t)$. On
the one hand, if $Vt \ll 1$, the integral becomes $\cosh(L) \int_0^t \alpha \dd
s/ (\alpha \tau_1 + s)$, leading to~:

\beq 
\tau(r) = \alpha \tau_1 \left[ \exp \left( -\frac{\ln(r)}{\alpha \ \cosh(L)} 
\right)-1 \right].
\eeq

On the other hand, $Vt \ge 1$ describes situations where the force $k l(t)$
approaches the critical depinning value. It's very unlikely to find high value
$L \gg 1$ or negative $ L \ll -1$. In order to cover the range $Vt \simeq 1$,
$\Pi^{-1}$ has been tabulated, by a direct numerical calculation, for a grid in
the plane $(L,\ln V)$ with variable $L$ between $-3<L<5 \ ; \ \Delta L =0.1$
and variable $V$, $10^{-8}<V<1$ with 20 values following a geometrical
sequence. At each point of the grid $\Pi^{-1}(r)$ is approximated with a
Tchebitchev polynomials approximation, by keeping  the 30 first coefficients.

When a couple $(L,\ln V)$ not on the grid is required, its  Tchebitchev
coefficients are linearly interpolated. The conversion $\tau(r)$
is reduced to a fast polynomial evaluation.

The two previous approximations match fine for \\ \hbox{$Vt \simeq 0.01$}, and were
actually used in the simulation.



\end{multicols}

\clearpage

\begin{thebibliography}{10}

\bibitem{FeiVin}
M. {Feigel'man} and V. Vinokur, Journal de Physique I(France) {\bf 49},  1731
  (1988).

\bibitem{Bou}
J.P. Bouchaud, Journal de Physique {\bf 2},  1705  (1992).

\bibitem{May}
R. Maynard, J. de Physique-Lettres {\bf 45},  L81  (1984).

\bibitem{Sch}
S. Scheidl, Zeitschrift f{\"u}r physik B {\bf 97},  345  (1995).

\bibitem{LeDVin}
P. Le~Doussal and V. Vinokur, Physica C {\bf 254},  63  (1995).

\bibitem{rem:toymodel} S. Franz and M. M\'ezard, EuroPhys.Lett. {\bf 26}, 209 
(1994). \\
L. Cugliandolo and P. Le~Doussal, Phys.Rev.E  {\bf 53},  1525  (1996).

\bibitem{rem:entropic} See for example 
J. Kurchan and L. Laloux, J. of physics A {\bf 29},  1929  (1996), \\
F. Ritort Phys.Rev.Lett {\bf 75}, 1190 (1995),\\
A. Barrat and M. M\'ezard J.de Physique (France) {\bf 5}, 941 (1995).

\bibitem{remarque2} By adjusting the distribution of the successive barriers
that a particle, in a one dimensional case, has to overcome, L. Laloux and P.
Le~Doussal are able to reproduce almost all the possible ageing behaviors.
cond-mat/{\bf 9705249}.

\bibitem{BlaGesFeiLarVin}
G. Blatter {\it et~al.}, Review of Modern Physics {\bf 66},  1125  (1994).

\bibitem{BouGeo}
J.P. Bouchaud and A. Georges, Physics Reports {\bf 195},  127  (1987).

\bibitem{HesBauPerCarCar}
F. Heslot {\it et~al.}, Phys.Rev.E {\bf 49},  4973  (1994).

\bibitem{CarNoz}
C. Caroli and P. Nozi\`eres,  in {\em Physics of Sliding Friction} (Kluwer
  Academic, Dordrecht, 1996).

\bibitem{HanTalBor}
P. H{\"a}nggi, P. Talkner, and M. Borkovec, Review of Modern Physics {\bf 62},
  251  (1990).

\bibitem{BorKalLeb}
A. Bortz, M. Kalos, and J. Lebowitz, J.Comput.Phys {\bf 17},  10  (1975).

\bibitem{Hor}
H. Horner, Zeitschrift f{\"u}r physik B {\bf 100},  243  (1996).

\bibitem{Fis:3} D. Fisher, Phys.Rev.Letters {\bf 50},  1486  (1983). For the
depinning of  Charge Density Waves.

\bibitem{reviewmovingphase}
For a review of velocity induced phase transitions in the context of high Tc
superconductors see
A. Koshelev and V. Vinokur, Phys.Rev.Lett {\bf 73},  3580  (1994).\\
T. Giamarchi and P. Le~Doussal, preprint cond-mat/ {\bf 9708085},    (1996).

\end{thebibliography}
\end{document}